\documentclass[prl,superscriptaddress,showpacs,twocolumn]{revtex4}

\usepackage{graphicx}

\usepackage{amssymb,amsmath}
\usepackage[nooneline,raggedright]{subfigure}

\usepackage{multirow}

\usepackage[dvips]{color}
\newcommand{\rowcolor}[1]{}
\newcommand{\cellcolor}[1]{}

\newcommand{\eref}[1]{Eq.~(\ref{#1})}
\newcommand{\fref}[1]{Fig.~\ref{#1}}

\newcommand{\tref}[1]{Tab.~\ref{#1}}

\definecolor{gray}{rgb}{0.7,0.7,0.7}        
\definecolor{grey}{rgb}{0.7,0.7,0.7}        
                
\newcommand{\bra}[1]{\ensuremath{\langle #1|}}
\newcommand{\ket}[1]{\ensuremath{|#1\rangle}}

\newcommand{\vek}[1]{%
        \hbox{\textbf #1}}

\newcommand{\svek}[1]{%
        {\mathbf #1}}

\newcommand{\out}[1]{}

\newcommand{\pr}{%
        ^\prime}

\begin{document}

\title{Many-body effects in iron pnictides and chalcogenides --\\ 
non-local vs dynamic origin of effective masses}

\author{Jan M. Tomczak}
\affiliation{Department of Physics and Astronomy, Rutgers University, Piscataway, New Jersey 08854, USA}
\author{M. van Schilfgaarde}
\affiliation{Department of Physics, King's College London, Strand, London WC2R 2LS}
\author{G. Kotliar}
\affiliation{Department of Physics and Astronomy, Rutgers University, Piscataway, New Jersey 08854, USA}

\begin{abstract}
We apply the quasi-particle self-consistent {\it GW} (QS{\it GW}) approximation to
some of the iron pnictide and chalcogenide superconductors.
We compute Fermi surfaces and density of states,
and
find excellent agreement with experiment,
substantially improving over standard band-structure methods.
Analyzing the QS{\it GW} self-energy we discuss non-local and dynamic contributions to effective masses.
We present evidence that the two contributions are mostly separable,
since the quasi-particle weight is found to be essentially independent of momentum.
The main effect of non locality is captured by the static but non-local QS{\it GW} effective potential.
Moreover, these non-local self-energy corrections, absent in e.g.\
dynamical mean field theory (DMFT), can be relatively large.
We show, on the other hand, that QS{\it GW} only partially accounts for
dynamic renormalizations at low energies.
These findings suggest that QS{\it GW} combined with DMFT will capture most
of the many-body physics in the iron pnictides and chalcogenides. 
\end{abstract}

\maketitle

\noindent
The discovery of superconductivity in the iron pnictides and chalcogenides has triggered much effort into understanding
their electronic properties.\cite{pnictide_review} 
The first theoretical insight into the pnictides was gained by
density functional theory (DFT) within the local density approximation (LDA), which 
correctly predicted the Fermi surfaces of LaFePO\cite{PhysRevB.75.035110} and LaFeAsO\cite{PhysRevLett.100.237003},
as well as the striped antiferromagnetic spin ground state\cite{0295-5075-83-2-27006}.
However there is ample experimental evidence for sizable correlation effects in the pnictides,
manifesting themselves
in high effective masses as witnessed by optical spectroscopy\cite{Qazilbash_pnictide},
de Haas--van Alphen measurements\cite{JPSJ.79.053702},  or bandwidth renormalizations from photoemission spectroscopy (PES) \cite{2010arXiv1009.0271W,PhysRevB.80.174510,Brouet2011_bfa_arx,PhysRevLett.105.067002,PhysRevB.82.184511},
as well as low magnetic moments 
\cite{PhysRevLett.101.257003,PhysRevB.82.064409}.
These issues have been successfully addressed by 
 many-body techniques,
yielding correct effective masses\cite{Yin_pnictide,PhysRevB.82.064504}, ordered moments\cite{Yin_pnictide},
and good structures\cite{PhysRevB.84.054529,PhysRevLett.104.047002}.

In many of these works, however, correlation effects 
have been accounted for by treating {local} interactions 
for a subspace of orbitals only. 
While this improves the description, several shortcomings persist~:
(a) correcting only a subspace means that large parts of the electronic structure
are still DFT derived. 
The choice of the exchange-correlation functional being discretionary, 
moreover causes
a dependence  
on the effective one-particle starting point.
Also, results depend on the correlated subspace chosen.
(b) Out-of-subspace self-energies\cite{ferdi_down} are neglected,
leading e.g.\ to underestimates in p-d gaps.
(c) Non local interactions are treated on the effective one-particle level.
To address these issues, 
we applied the quasi-particle self-consistent {\it GW} approximation
\cite{PhysRevLett.93.126406,schilfgaarde:226402,fpgw} 
to the iron pnictides and chalcogenides. 
We discuss band-structures, Fermi surfaces, and density of states (DOS) and elucidate the origin of many-body effective masses.
%
The essence of our findings is~:
(1) Non-local self-energy effects are not small. These are neglected in DMFT\cite{RevModPhys.78.865} 
based approaches.
(2) QS{\it GW}, which includes non-local correlations, produces excellent Fermi surfaces, but
(3) does not adequately account for the dynamics of the self-energy.
(4) Non-locality and energy dependence are shown to be essentially separable.
(5) Together, this indicates that a combined QS{\it GW}+DMFT approach is a very promising avenue in electronic structure theory.

An  important goal of the many-body theory of solids is to capture the one-particle Greens-function $G$, which is written as 
$G^{-1}=\omega-H^0(\svek{k})-\Sigma(\svek{k},\omega)$, with $H^0$ a reference one-particle Hamiltonian, and $\Sigma$ a self-energy 
that is defined with respect to correlation effects already included in $H^0$\footnote{
A matrix structure in the space of orbitals is assumed.}. 
For example, in LDA+DMFT\cite{RevModPhys.78.865}, $H^0$$=$$H^{\hbox{\tiny LDA}}$$=$$-\nabla^2+v_{crystal}+v_{Hartree}+v_{xc}^{\hbox{\tiny LDA}}$, is the Kohn-Sham Hamiltonian,
with $\Sigma$$=$$(\Sigma^{\hbox{\tiny DMFT}}(\omega)-E_{dc})_{LL\pr RR\pr}\ket{RL}\bra{R\pr L\pr}$.
The self-energy $\Sigma^{\hbox{\tiny DMFT}}$ acts on a set of correlated orbitals $\ket{RL}$,
and the portion $E_{dc}$ of $\Sigma^{\hbox{\tiny DMFT}}$ already contained in $H^0$ must be
subtracted.
In the QS{\it GW} approximation, $H^0$$=$$H^{\hbox{\tiny QS{\it GW}}}$$=$$-\nabla^2$$+$$v_{crystal}$$+$$v_{Hartree}$$+$$v_{xc}^{\hbox{\tiny QS{\it GW}}}$, and $\Sigma^{\hbox{\tiny QS{\it GW}}}$$=$$G^{\hbox{\tiny QS{\it GW}}}W-v_{xc}^{\hbox{\tiny QS{\it GW}}}$, 
where the (static) QS{\it GW} exchange-correlation potential, $v_{xc}^{\hbox{\tiny QS{\it GW}}}$, is chosen so as to closely mimic the same quasi-particles $E_{\svek{k}j}$ in $H^{\hbox{\tiny QS{\it GW}}}$ as in $(\omega-H^{\hbox{\tiny QS{\it GW}}}-\Sigma^{\hbox{\tiny QS{\it GW}}})$.
These are given by 
\begin{equation}
\left[ H^{\hbox{\tiny QS{\it GW}}}(\svek{k})+\Re\Sigma^{\hbox{\tiny QS{\it GW}}}(\svek{k},E_{\svek{k}j})\right]\ket{\Psi_{\svek{k}j}}=E_{\svek{k}j}\ket{\Psi_{\svek{k}j}}
\label{qs}
\end{equation}
and  it
can be shown\cite{PhysRevLett.93.126406} 
that a good choice for $v_{xc}^{\hbox{\tiny QS{\it GW}}}$ is
\begin{equation}
\frac{1}{2}\sum_{ij\svek{k}}\ket{\Psi_{\svek{k}i}}\Re\left[\Sigma_{ij}^{\hbox{\tiny QS{\it GW}}}(\svek{k},E_{\svek{k}i})+\Sigma_{ji}^{\hbox{\tiny QS{\it GW}}}(\svek{k},E_{\svek{k}j})  \right]\bra{\Psi_{\svek{k}j}}.
\label{qsGW}
\end{equation}

\begin{figure*}[htp]    
\centering
\subfigure[\,]{\scalebox{.8}{\includegraphics[clip=true,trim=17 15 2 0, angle=-90,width=0.545\textwidth]{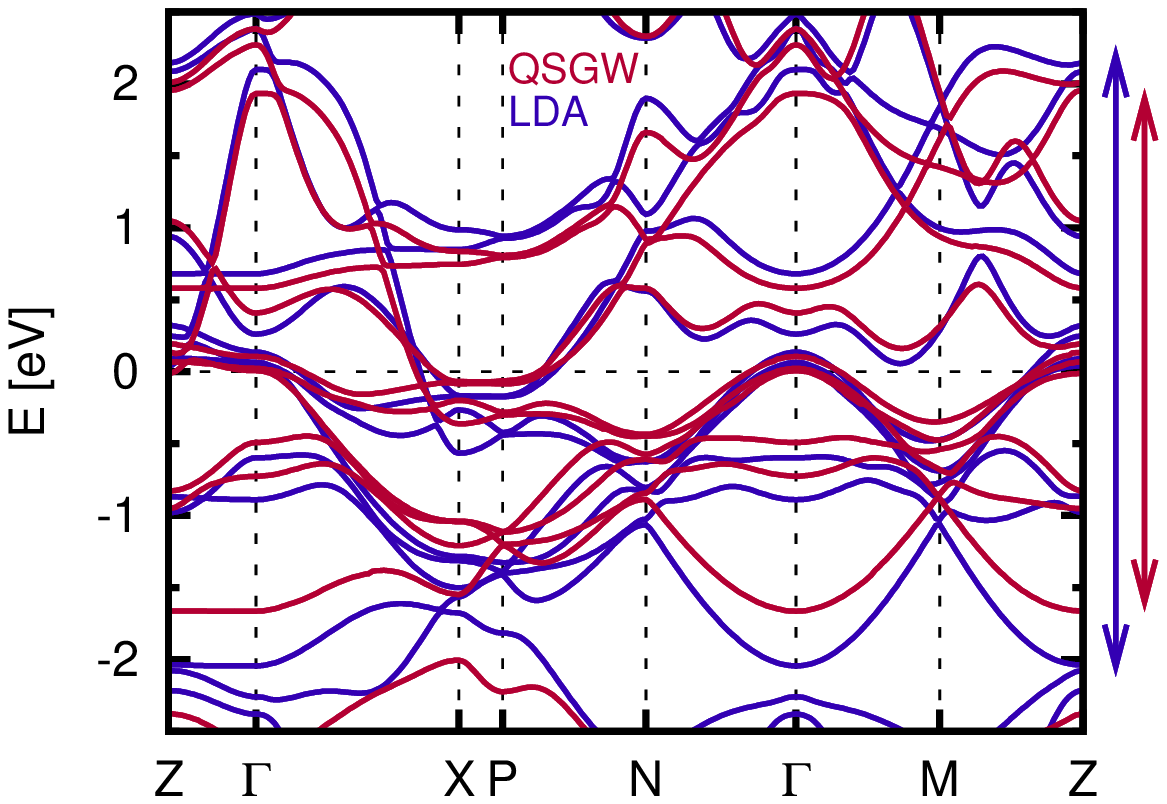}}}
\hspace{1cm}
\subfigure[\,]{\scalebox{.875}{\includegraphics[clip=true,trim=17 5 12 25, angle=-90,width=0.34\textwidth]{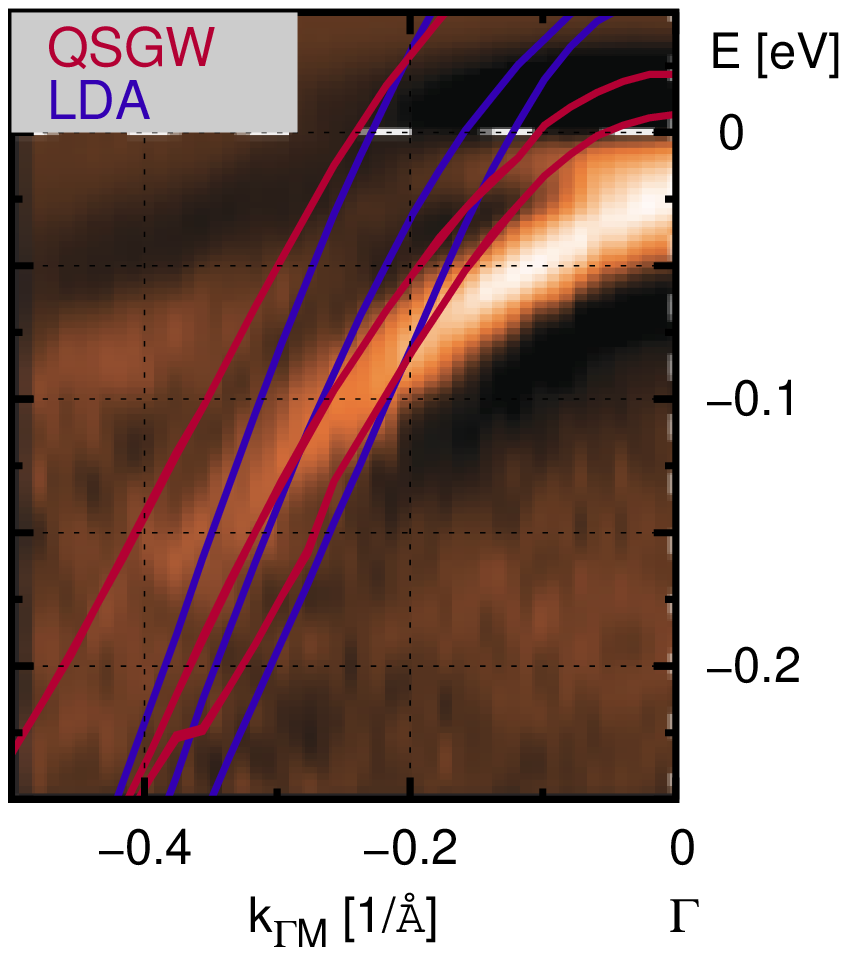}}}
\caption{{\bf paramagnetic BaFe$_2$As$_2$.} (a) bandstructures within LDA and QS{\it GW} (notice the Fe-3d bandwidth narrowing as indicated by the arrows), (b) comparison with ARPES\cite{PhysRevB.83.054510} around the $\Gamma$ point. 
The slight constant energy shift between theory and experiment possibly originates from doping in the measured sample,
BaFe$_{1.85}$Co$_{0.15}$As$_2$\cite{PhysRevB.83.054510}.
}
\label{fig1}
\end{figure*}

Results for paramagnetic BaFe$_2$As$_2$ are shown in \fref{fig1}. 
As is clearly visible, QS{\it GW} substantially narrows the iron 3d bands
(by 16\%) relative to the Kohn-Sham spectrum within LDA%
\footnote{Our LDA results (method of Ref.~\onlinecite{fplmto}) are
congruent with previous works, e.g.\ Ref.~\onlinecite{JPSJ.79.044705}.
}%
. 
This narrowing corresponds to an enhancement of the effective mass, $m^{\hbox{\tiny QS{\it GW}}}/m^{\hbox{\tiny LDA}}$, and is slightly larger
than the 12\% found for elemental iron\cite{schilfgaarde:226402}
(See \tref{tab1} for other compounds). 
In \fref{fig1}(b) 
we compare QS{\it GW} and LDA bands near the Fermi level $E_F$ and $\vek{k}$ near zero,
to angle resolved photoemission (ARPES) experiments\cite{PhysRevB.83.054510}.
It is consensus from experimental data that there is,
 at the $\Gamma$ point, an outer hole pocket, while closer to $\Gamma$
there are two xz/yz excitations 
\cite{PhysRevB.81.104512,PhysRevB.83.054510,Brouet2011_bfa_arx}. 
The top of these inner bands is very close to zero\cite{PhysRevB.83.054510}, particularly for Co-doped samples\cite{PhysRevB.83.054510,JPSJ.80.113707}.
The character of the outer pocket is debated~: a mixture of xy and x$^2$-y$^2$ \cite{PhysRevB.83.054510},
z$^2$ additions\cite{PhysRevB.83.064516},
and a change of character along $k_z$\cite{PhysRevB.83.064516} is reported.
The Fermi vector has been measured
for the outer band only~: 
$k_F$$=$$0.07$$-$$0.17$\AA$^{-1}$\cite{PhysRevB.81.104512,PhysRevB.79.155118,PhysRevB.83.054510}.
QS{\it GW} hardly effects  
this pocket,
and it indeed is mainly of xy character 
for both LDA and QS{\it GW}.
The Fermi wave vector is 
$k_F$$=$$0.24$\AA$^{-1}$. 
The inner pockets are of xz/yz character, in accord with experiment
\cite{PhysRevB.83.054510,JPSJ.80.113707}.
Relative to LDA,
their size is reduced to
$k_F$$=$$0.05$\AA$^{-1}$ 
and $k_F$$=$$0.11$\AA$^{-1}$%
\footnote{Interestingly, the spin-orbit coupling makes a qualitative difference here.
Indeed without it, the pockets are reversed in LDA, thus yielding wrong orbital characters.
}.

While QS{\it GW} reduces the group velocity relative to LDA, 
the dispersion at $k_F$ near $\Gamma$ as measured by ARPES is still a factor of
two or so smaller.
This suggests that QS{\it GW} does not fully account for all
many-body renormalizations in this compound, as we detail below.

\begin{figure*}[htp]
\centering
\subfigure[\, band-structure of FeSe]{\scalebox{.8}{\includegraphics[clip=true,trim= 13 0 2.5 0, angle=-90,width=0.45125\textwidth]{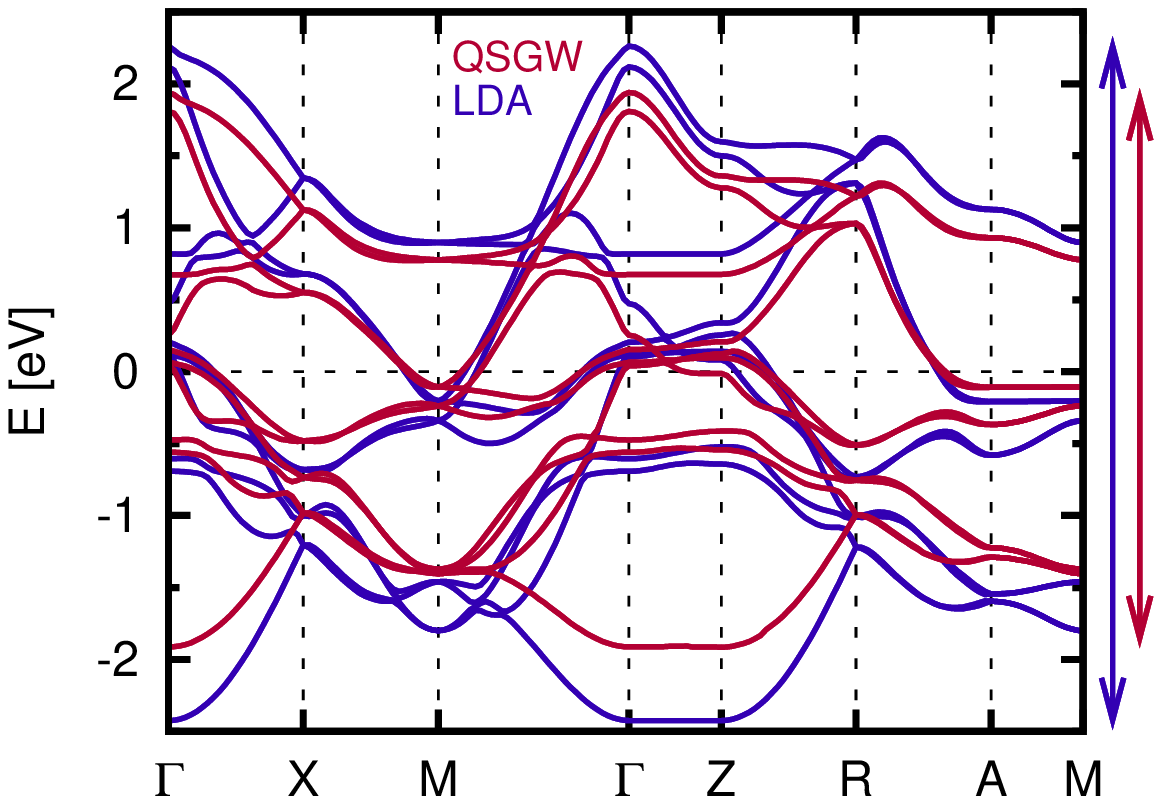}}}
\subfigure[\, FeSe density of states vs. photoemission]{\scalebox{.8}{\includegraphics[clip=true,trim= 13 0 2.5 0,angle=-90,width=0.45125\textwidth]{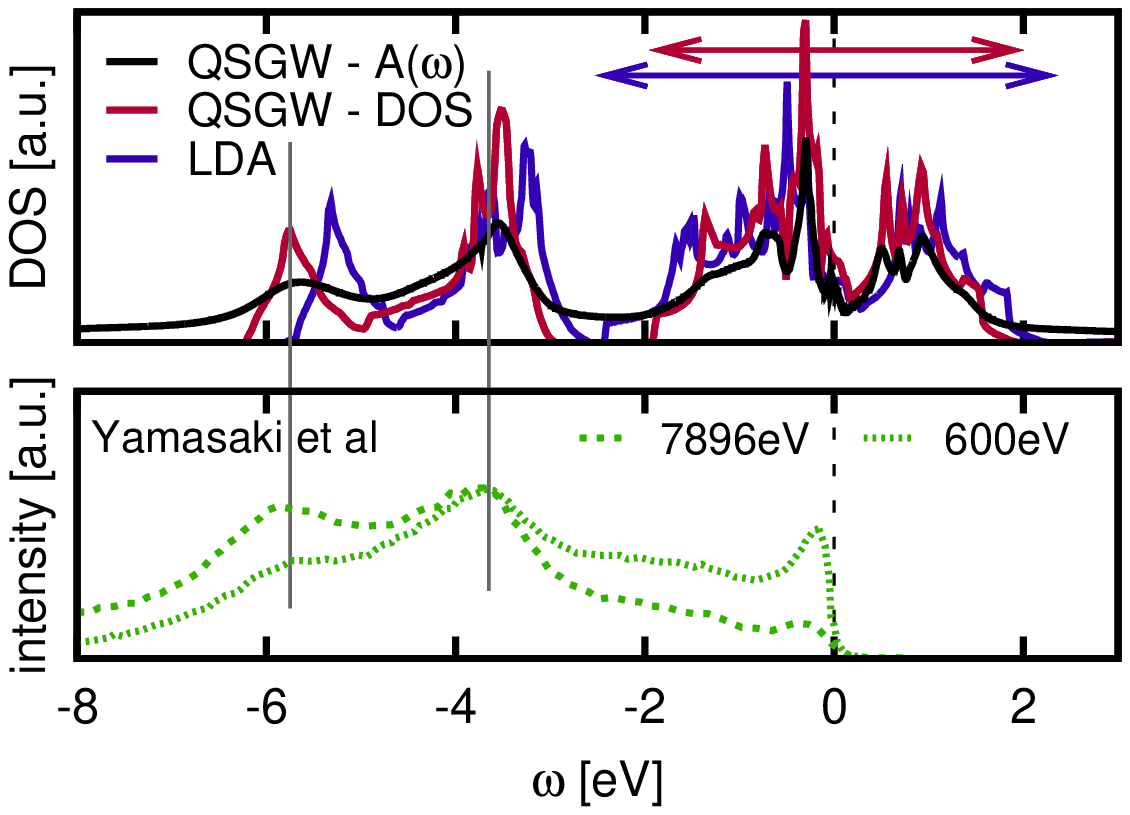}}}\\
\subfigure[\, FeSe]{\scalebox{.75}{\includegraphics[clip=true,trim=40 70 33 60,angle=-90,width=0.3015\textwidth]{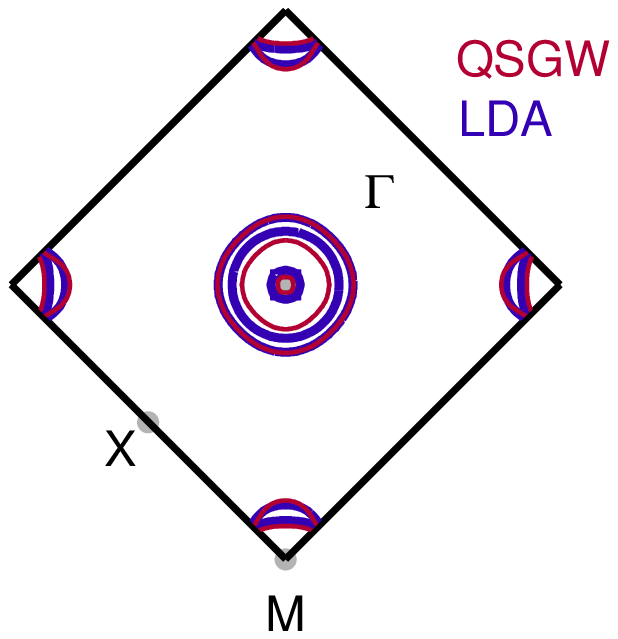}}}
\subfigure[\, Fe$_{1.04}$Te$_{0.66}$Se$_{0.34}$]
                   {\scalebox{.75}{\includegraphics[clip=true,trim=12 0 47 0,angle=-90,width=0.27\textwidth]{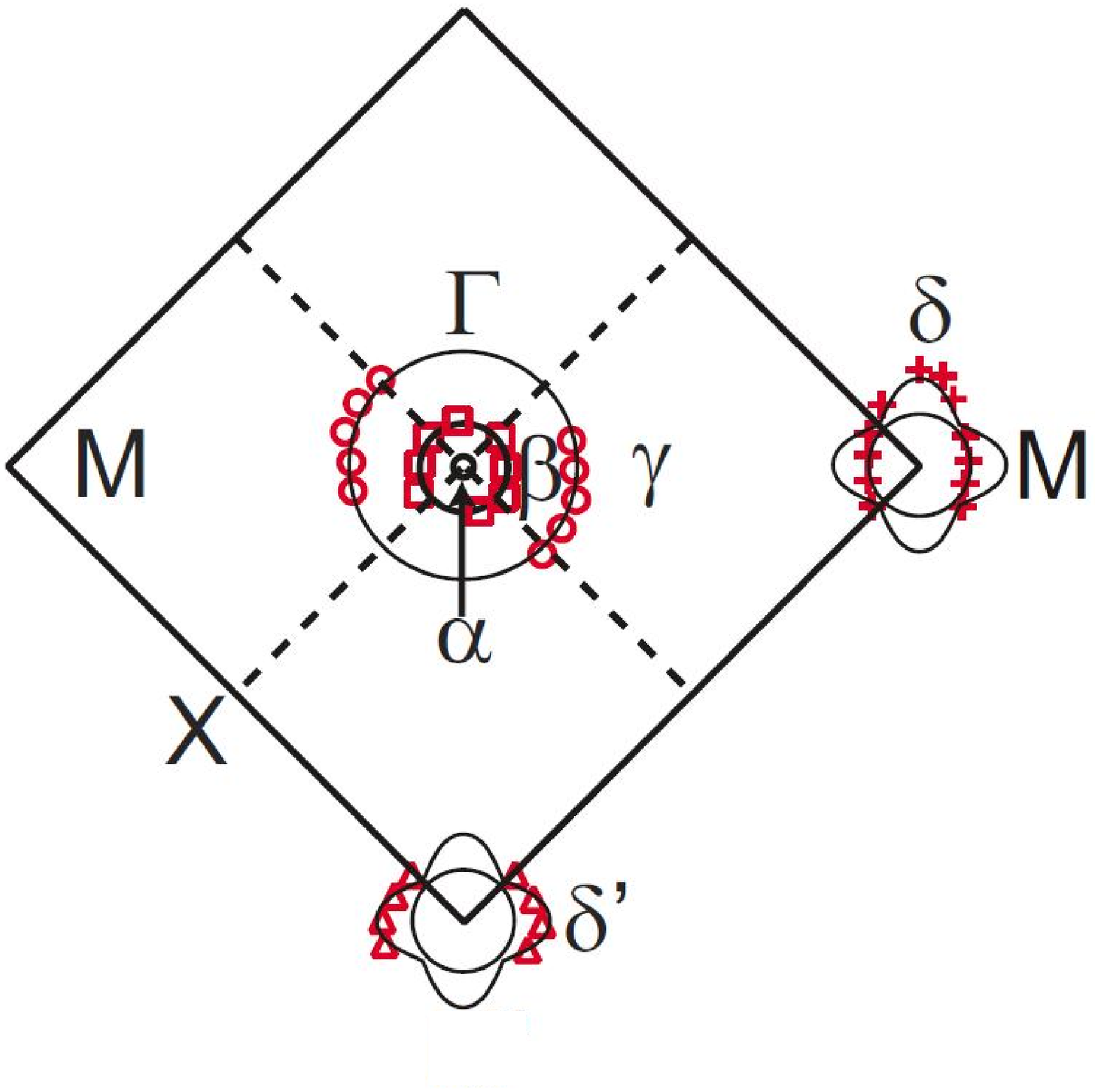}}}
\subfigure[\, FeTe]{\scalebox{.75}{\includegraphics[clip=true,trim=40 60 33 70,angle=-90,width=0.3015\textwidth]{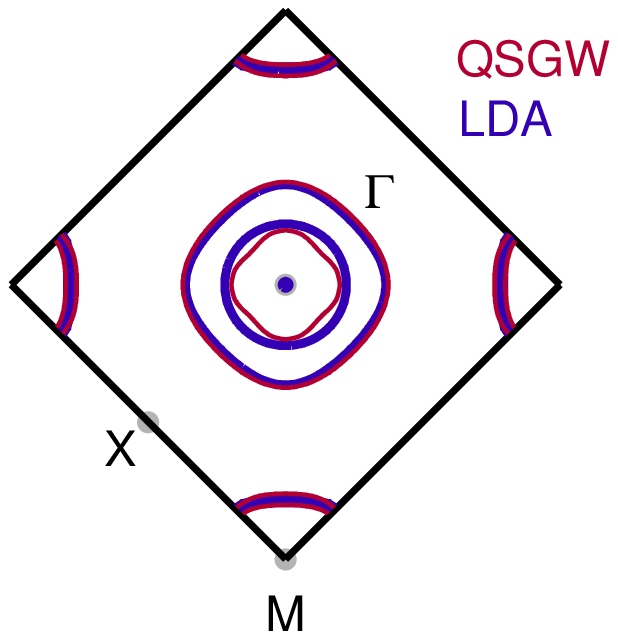}}}
\caption{{\bf paramagnetic FeSe.} (a) band-structure within LDA and QS{\it GW} (notice the substantial bandwidth narrowing indicated by the arrows), (b) DOS in comparison to photoemission\cite{PhysRevB.82.184511}.
Also shown is the spectral function $A(\omega)$ of the QS{\it GW} that takes into account lifetime effects.
Arrows indicate the Fe-3d bandwidth; grey lines are guides to the eye.
(c)-(e) Fermi surfaces~: (c)/(e) FeSe/FeTe within QS{\it GW} and LDA, (d) experimental Fermi surface of Fe$_{1.04}$Te$_{0.66}$Se$_{0.34}$ from ARPES\cite{PhysRevB.81.014526}.}
\label{fig2}
\end{figure*}

We next consider the chalcogenide FeSe (\fref{fig2}).
The QS{\it GW} band-structure displays a 
remarkable narrowing of the iron 3d bands of 22\%, equally visible in the DOS. 
Also higher energy bands get renormalized~: Se-4p excitations
are pushed down in energy, notably improving the agreement with PES.\cite{PhysRevB.82.184511}
In \fref{fig2}(c-e) we compare the Fermi surfaces of FeSe and FeTe within QS{\it GW} and LDA with 
measurements of the Fe$_{1.04}$Te$_{0.66}$Se$_{0.34}$ alloy\cite{PhysRevB.81.014526}.
In experiment 
there are three hole pockets 
at $\Gamma$ and two electron pockets around $M$.
QS{\it GW} and LDA predict similar surfaces for the xy pocket labeled ``$\gamma$'',
 and the Fermi wave vectors (FeSe : $k_F^\gamma=0.29$\AA$^{-1}$, FeTe : $k_F^\gamma=0.43$\AA$^{-1}$) encompass the ARPES value for Fe$_{1.04}$Te$_{0.66}$Se$_{0.34}$ 
 ($k_F^\gamma=0.3$\AA$^{-1}$).
The QS{\it GW} Fermi vectors of the xz/yz ``$\alpha$'' and ``$\beta$'' pockets shrink relative
to the LDA, and become more anisotropic. Indeed, the ``$\alpha$'' pocket has 
$k_F^\alpha=0.04$\AA$^{-1}$ on the $\Gamma M$ line for
FeSe, and completely disappears in FeTe.
The magnitude agrees well with ARPES measurements ($k_F^\alpha=0.03$\AA$^{-1}$),
whereas the ARPES $\beta$ pocket ($k_F^\beta=0.12$\AA$^{-1}$)
is somewhat smaller than in QS{\it GW} ($k_F^\beta=0.19$\AA$^{-1}$ and $0.23$\AA$^{-1}$ in FeSe and FeTe,
respectively).

\begin{figure}[!b!]
\scalebox{.8}{\includegraphics[clip=true,trim=17.5 70 17.25 70,angle=-90,width=\columnwidth]{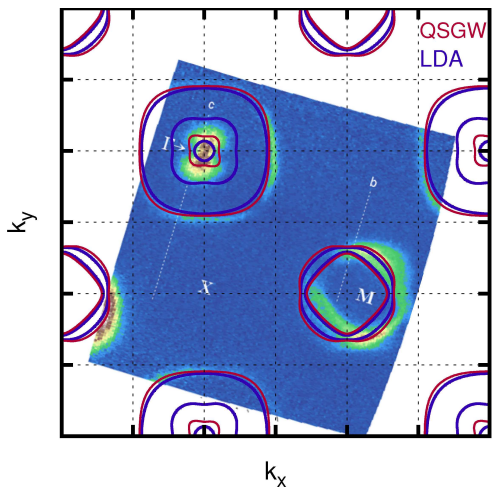}}
\caption{{\bf Fermi surface of LiFeAs.} $k_z=0$--plane in the Brillouin zone for 2 Fe atoms; experimental intensity from Ref.~\onlinecite{PhysRevLett.105.067002}.
Notice how QS{\it GW} drastically shrinks the inner pockets at $\Gamma$.
}%
\label{fig3}
\end{figure}

For the 111-family, we show in \fref{fig3} the Fermi surface of LiFeAs within
QS{\it GW}, LDA and ARPES\cite{PhysRevLett.105.067002}.
QS{\it GW} and ARPES agree very well for both large and small pockets:
note in particular how QSGW shrinks the inner pockets at $\Gamma$.
Both calculations and experiment\cite{PhysRevB.85.094509} concur in the orbital characters along $\overline{\Gamma X}$~:
xy, yz, and xz, for the outer, middle and inner band, respectively.
Within QS{\it GW}, the latter has moved below $E_F$.
At $M$ the smaller electron pocket is of mixed xz/yz; the bigger one, of
xy character, is too small in LDA.
It becomes larger in QS{\it GW}, as was previously found within LDA+DMFT\cite{Yin_pnictide,PhysRevB.85.094505}.

We turn to
the mass enhancement
relative to the LDA band masses, which have been used as a reference in the analysis of experiments.
The enhancement is given as the ratio of the magnitude of the LDA and the QS{\it GW}
group velocities near the Fermi level. 
The latter are 
\begin{equation}
\frac{dE_{\svek{k}i}}{dk_\alpha}=\frac{  \bra{\Psi_{\svek{k}i}}   \partial_{k_\alpha} \left( H^{\hbox{\tiny QS{\it GW}}}+\Re\Sigma^{\hbox{\tiny QS{\it GW}}}(\omega=0)\right)
\ket{\Psi_{\svek{k}i}} } {\left( 1- \bra{\Psi_{\svek{k}i}} \partial_\omega\Re\Sigma^{\hbox{\tiny QS{\it GW}}}\ket{\Psi_{\svek{k}i}}\right)^{-1}_{\omega=0}}
\label{landau}
\end{equation}
\begin{table}[t]%
\begin{tabular}{l|c|cc|cc|cc} 
&  QS{\it GW}  & \multicolumn{2}{c|}{QS{\it GW}}  &  \multicolumn{2}{c|}{ARPES}  &\multicolumn{2}{c}{DMFT}  \\
& \multirow{2}{*}{$\frac{m^{\hbox{\tiny QS{\it GW}}}}{m^{\hbox{\tiny LDA}}}$}
 & \multicolumn{2}{c|}{$1/Z^{\hbox{\tiny QS{\it GW}}}$ } &  \multicolumn{2}{c|}{$m^*/m^{\hbox{\tiny LDA}}$}    & \multicolumn{2}{c}{$1/Z^{\hbox{\tiny DMFT}}$} \\
& & \phantom{/y}xy\phantom{x} & xz/yx  &  xy & xz/yz\\
\hline
CaFe$_2$As$_2$ & 1.05 & 2.2 & 2.1 &   \multicolumn{2}{c|}{2.5\cite{2010arXiv1009.0271W}}  & 2.7 & 2.0 \\ 
SrFe$_2$As$_2$ & 1.13 & 2.3 & 2.0 &   \multicolumn{2}{c|}{3.0\cite{PhysRevB.80.174510}}   &  2.7 & 2.6 \\
BaFe$_2$As$_2$ & 1.16 & 2.2 & 2.2 &  2.7  & 2.3 \cite{Brouet2011_bfa_arx} &  3.0 & 2.8  \\  
LiFeAs         & 1.15 & 2.4 & 2.1 &   \multicolumn{2}{c|}{3.0\cite{PhysRevLett.105.067002}} &     3.3/2.8   & 2.8/2.4 \\
FeSe           &  1.22 & 2.4   & 2.2 & \multicolumn{2}{c|}{3.6\cite{PhysRevB.82.184511}} &  3.5/5.0  &  2.9/4.0 \\ 
FeTe           &  1.17 & 2.6  & 2.3   &  \multicolumn{2}{c|}{6.9 \cite{PhysRevLett.104.097002}}& 7.2    &  4.8  
\end{tabular}
\caption{{\bf effective masses and quasi-particle weights.} 
$m^{\hbox{\tiny QS{\it GW}}}/m^{\hbox{\tiny LDA}}$ denotes the ratio of the iron 3d bandwidth within QS{\it GW} and LDA. The
quasi-particle weights $Z^{\hbox{\tiny QS{\it GW}}}$ are extracted from $\Sigma^{\hbox{\tiny QS{\it GW}}}$ 
at the $\Gamma$ point in the band basis.
DMFT masses from Ref.~\onlinecite{Yin_pnictide}, except for the second values of LiFeAs (Ref.~\onlinecite{PhysRevB.85.094505}) and FeSe (Ref.~\onlinecite{PhysRevB.82.064504}).
ARPES effective masses are obtained with respect to LDA, in accordance with \eref{landau}.
\label{tab1}}
\end{table}
A change in the velocity, and thus the effective mass, is possible through (a) the {\it dynamical} part of the self-energy 
via the quasi particle weight $Z_k$ $=$ $1/\left(1-
\partial_\omega\Re\Sigma^{\hbox{\tiny QS{\it GW}}}(k,\omega)\right)_{\omega=0}$.
Noting that $H^{\hbox{\tiny QS{\it GW}}}+\Sigma^{\hbox{\tiny QS{\it GW}}}$$=$$-\nabla^2+v_{crystal}+v_{Hartree}-G^{\hbox{\tiny QS{\it GW}}}W$ there is (b)
a renormalization of the velocity through {\it non-local} correlations as encoded in $G^{\hbox{\tiny QS{\it GW}}}W$,
and (c) an effect through a change in charge density.

In DMFT approaches, the self-energy is local by construction, and enhanced masses emerge solely through the {\it energy
dependence} of the self-energy (indeed $m/m^{\hbox{\tiny DMFT}}$$=$$Z^{\hbox{\tiny DMFT}}$).
While in many correlated materials, in which the physics is controlled by the on-site Coulomb interaction,
there is evidence that this is a good approximation\cite{vollkot}, it is not clear {\it a priori} whether such an ansatz is warranted for the pnictides and chalcogenides.
QS{\it GW}, on the other hand, accounts, albeit perturbatively, for all mechanisms of effective masses~: Indeed, the momentum-dependence 
is manifestly included via \eref{qsGW}, and the energy slope enters through the 
QS{\it GW} procedure, \eref{qs},
that determines $E_\svek{k}$. 
By construction, \eref{landau} also reads
$\frac{dE_{\svek{k}i}}{dk_\alpha}=  \bra{\Psi_{\svek{k}i}}  \left( \partial_{k_\alpha}  H^{\hbox{\tiny QS{\it GW}}} \right) 
\ket{\Psi_{\svek{k}i}} $, hence $v_{xc}^{\hbox{\tiny QS{\it GW}}}$ accounts for, both,
the dynamic and non-local  renormalizations%
\footnote{For insulating VO$_2$ 
correlation effects from a DMFT cluster extension were found to be mappable onto a non-local one-particle potential\cite{me_vo2}
akin to the QS{\it GW} procedure.}%
.

In order to discuss non-local and dynamic contributions to effective masses, we 
analyze the QS{\it GW} self-energy~: 
\tref{tab1} summarizes our results for
the dynamical renormalization factors $1/Z^{\hbox{\tiny QS{\it GW}}}$, the net
3d bandwidth-narrowing, $m^{\hbox{\tiny QS{\it GW}}}/m^{\hbox{\tiny LDA}}$\footnote{%
We find the ratio of bandwidths is very similar to the inverse ratio of group velocities on the Fermi surface.
}%
, as well as results from DMFT and ARPES.
This reveals~:
(a) 
The 
inverse quasi-particle weights
are larger than the actual bandwidth narrowing by a
factor of two or more.
One clear source of the discrepancy originates in the nonlocality of the self-energy, missing in both LDA and DMFT.
Nonlocality tends to delocalize quasi-particles\cite{PhysRevB.71.045323}, see \eref{landau},
which partially cancels the effects of the energy dependence of $\Sigma$.
We analyze this further below.
(b) The band-width narrowing $m^{\hbox{\tiny QS{\it GW}}}/m^{\hbox{\tiny LDA}}$ is too small with respect to experiment. 
This is generally expected, since the reduction of the quasi-particle weight $Z$
in correlated systems is a largely non-perturbative effect -- a realm where DMFT excels.
In particular, a DMFT study\cite{Yin_pnictide} rationalized that the large effective mass in FeTe is 
owing to the local Hund's rule coupling, a multiplet effect not well treated in {\it GW}.
Despite the insufficient dynamic renormalization, however, the 
trends in $Z$ along the series of materials are captured, including the orbital differentiation. 
Indeed hole-like excitations of xy character are heavier than xz/yz ones for all compounds considered.

To make a direct connection with DMFT, we introduce a local basis set (indexed $L$) by constructing
maximally localized Wannier functions\cite{PhysRevB.56.12847,miyake:085122} for the iron 3d and pnictogen/chalcogen 4p orbitals.
We further introduce the momentum variance $(\delta_k X)^2$$=$$1/{N_L}^2\sum_{\svek{k}LL\pr}\left|  \Re X_{LL\pr}^{\svek{k}}- \Re{X}_{LL\pr}^{loc}  \right|^2$
of a quantity $X$ with respect to its local reference ${X}_{LL\pr}^{loc}$$=$$\sum_{\svek{k}}X_{LL\pr}^{\svek{k}}$\footnote{We restrict $L$, $L\pr$ to the 3d orbitals of iron.
}.
To quantify
static corrections in QS{\it GW} beyond LDA at the Fermi level we define the many-body correction
$\widetilde{\Sigma}=\Sigma^{\hbox{\tiny QS{\it GW}}}+H^{\hbox{\tiny QS{\it GW}}}-H^{\hbox{\tiny LDA}}$, and compute its variance
$\delta_k\widetilde{\Sigma}(\omega$$=$$0)$ 
\footnote{The Wannier functions based on the LDA
and the QS{\it GW} are of course different. However, as the Wannier spread differ by less than 3\%, we neglect this effect.}.
We find $\delta_k\widetilde{\Sigma}$$=$$0.08$eV for BaFe$_2$As$_2$, and $\delta_k\widetilde{\Sigma}$$=$$0.09$eV for FeSe,
while the variations of the LDA exchange-correlation potential, $\delta_kv_{xc}^{\hbox{\tiny LDA}}$, are 0.22eV and 0.2eV, respectively.
Hence, QS{\it GW} overall reduces the momentum dependence,
consistent with the observed bandwidth-narrowing, see \eref{landau}.
The relative change in the momentum variation, $\delta_k\widetilde{\Sigma}/\delta_kv_{xc}^{\hbox{\tiny LDA}}$, 
is important in both compounds~: 36\% for BaFe$_2$As$_2$ and 45\% in FeSe.
This shows that the assumption of a purely local self-energy correction ({\`a} la DMFT) 
is not fully warranted.

To quantify the momentum dependence of the quasi-particle dynamics, 
we introduce the variance $\Delta_k Z=\sqrt{\sum_\svek{kL} |Z_{LL}^{\svek{k}}-Z_{LL}^{loc}|^2}$
of the  Fermi liquid weight.
We find $\Delta_k Z$$\approx$$0.5\%$ for all {\it GW} calculations performed here%
\footnote{%
$\Delta_k \left[1-
\partial_\omega\Re\Sigma^{\hbox{\tiny QS{\it GW}}}(k,\omega)\right]^{-1} $$<$$Z/10$
for $|\omega|$$<$$2$eV~: the momentum variance of the quasi-particle dynamics is small.
}%
. 
This establishes that
non-local and dynamical correlation effects are essentially {\it separable}. Thus the self-energy
is describable by the ansatz $\widetilde{\Sigma}(\svek{k},\omega)=F(\svek{k})+(1-Z^{-1})\omega$
within the Fermi liquid regime, which encompasses (within QS{\it GW}) the 
range $|\omega|\lesssim2\hbox{eV}$.
The QS{\it GW} self-consistency 
further imposes~:
$F(\svek{k})=Z^{-1}E_\svek{k}-\epsilon^{\hbox{\tiny LDA}}_\svek{k}$.

To introduce non-local correlations to DMFT, or, reversely, to improve on local fluctuations in {\it GW}, a
{\it GW}+DMFT scheme has been proposed\cite{PhysRevLett.90.086402}. In the simplest of implementations local and non-local 
self-energies from DMFT and {\it GW}, respectively, are combined. 
This was rationalized further by noting that self-energy contributions {\it beyond GW} are mostly local\cite{PhysRevLett.96.226403},
thus amenable to DMFT.
Here, 
we showed that {\it even on the GW level}, the (low-energy) dynamics is essentially local, too.
Hence, non-local correlations can be accounted for by the {\it static} QS{\it GW} potential.
This 
has several advantages~:
Due to its complexity,  {\it GW}+DMFT
has, so far, only been employed non-selfconsistently\cite{PhysRevLett.90.086402,jmt_svo}.
This  
defies many of the method's virtues, e.g.\ it persists a starting point dependence, and issues of double counting.
Here, we advocate to circumvent this 
by imposing (quasi-particle)
self-consistency for the
{\it GW} only, i.e.\ introducing
a QS{\it GW}+DMFT scheme. The latter is manifestly void of any DFT dependence, and the double counting is well defined
\footnote{namely by the local projection of $G^{\hbox{\tiny QS{\it GW}}}W$.}.

For the iron pnictides and chalcogenides we 
conclude that the QS{\it GW} approach yields very good Fermi surfaces, as well as sizable corrections to spectral features at finite energies.
We further showed that non-local correlations are important, 
and can be incorporated through the
QS{\it GW} effective potential.
Moreover we showed that dynamical correlations are mostly local.  They
are not sufficiently accounted for in QS{\it GW},
especially when correlations are strong, as in the chalcogenides.
These findings suggest that QS{\it GW} and DMFT taken together will be
sufficient to incorporate most of the correlations in  
electronic
structure.

The support of  the DOE-CMSN grant ``Computational Design of Fe Based Superconductors''
is gratefully acknowledged.


\end{document}